\documentclass[%
 reprint,
groupedaddress,
showpacs,%preprintnumbers,
 amsmath,amssymb,
 aps,
 prd,
 longbibliography,
 lengthcheck,%
]{revtex4-1}

\usepackage{graphicx}% Include figure files
\usepackage{epsfig}% пакет для графіків епс
\usepackage{epstopdf}% пакет який конвертує eps графіки у pdf - графіки
\usepackage{dcolumn}% Align table columns on decimal point
\usepackage{bm}% bold math
\usepackage{float}

\begin{document}

\title{Antigravity: Spin-gravity coupling in action}

\author{Roman Plyatsko and Mykola Fenyk}

\affiliation{Pidstryhach Institute for Applied Problems in
Mechanics and Mathematics\\ (National Academy of Sciences of Ukraine), \\ 3-b Naukova Street,
Lviv, 79060, Ukraine}

\begin{abstract}
The typical motions of a spinning test particle in  Schwarzschild's background which show the strong repulsive action of the highly relativistic spin-gravity coupling are considered using the exact Mathisson-Papapetrou equations. An approximated approach to choice solutions of these equations which describe motions of the particle's proper center of mass is developed.

\end{abstract}

\pacs{04.20.-q, 95.30.Sf}
%\keywords{}

\maketitle

\section{ Introduction}

For over 50 years the effects of general relativity in strong gravitational fields of massive compact objects (Schwarzschild's and Kerr's black holes, neutron stars, quasars) are  the focus of many studies.
Relevant results are presented in the classical books on general 
relativity \cite{Misner, Chandra}. Also note that the recent historical registration of the gravitational waves is directly related to the interaction and merger of the two massive black holes \cite{Abb}. 

The existence of some strong gravitational fields is not only caused by the large masses. It is pointed out in \cite{Pirani, Aich-1, Aich-2, Curtis, Eath-1, Eath-2, Eath-3, Mash-1, Mash-2} that when an
ordinary (not so great) Schwarzschild mass is moving with the velocity close to the speed of light its gravitational field becomes much grater than the field of this mass at rest. It means that  in terms of the gravitoelectric 
field $E^{(i)}_{(k)}$ and the gravitomagnetic field $B^{(i)}_{(k)}$ which are determined in \cite{Thorne} some components of  $E^{(i)}_{(k)}$
and $B^{(i)}_{(k)}$ are proportional to $\gamma$ or $\gamma^2$ ($\gamma$ is the relativistic Lorentz factor). The values $E^{(i)}_{(k)}$ determine the tidal forces \cite{Mash-1, Mash-2} whereas the components $B^{(i)}_{(k)}$ act on a spinning test particle (similarly as the usual magnetic field acts on a rotating charge) according to the known Mathisson-Papapetrou (MP) equations \cite{Mathis, Papa}. It is shown in \cite{Pl82, Pl98, Pl01, Pl05, Pl10, Pl11, Pl12, Pl13, Pl15, Pl-Sp} that just the highly relativistic regime of  spinning particle motions in Schwarzschild's and Kerr's background reveals new features of the gravitational interaction. (When the velocity of a spinning particle is not very high the gravitational spin-orbit and spin-spin interactions were considered in \cite{Wald}.) It is important that depending on the correlation of signs of the spin and the particle's orbital velocity the spin-gravity coupling acts as a significant repulsive or attractive force.

The purpose of this paper is to present new results concerning different physical situations in Schwarzschild's background when a spinning test particle feels the strong repulsive action caused by the highly relativistic spin-gravity coupling. Note that the consideration of these antigravity effects may be useful in the context of the repulsive phenomenon in cosmology. For example, some corresponding results can be generalized for the Schwarzschild--de Sitter metric.

The paper is organized as follows. In Sec. II we develop the results of paper \cite{Pl12} concerning the properties of the highly relativistic circular orbits of a spinning particle in Schwarzschild's background in the case of the strong repulsive action of the spin-gravity coupling: the energy and angular momentum on these orbits are considered. Sections III and IV are devoted to the specific repulsive features of the noncircular highly relativistic trajectories of a spinning particle which begins to move with 
$r_g<r\leq 1.5 r_g$. In Sec. IV an approximated method of selection solutions of the exact MP equations which describe the motions of the particle's proper center of mass is elaborated and used in computer calculations. We conclude in Sec. V.

\section{Energy and angular momentum of a spinning particle on highly relativistic circular orbits in Schwarzschild's background}

It is known that the geodesic circular orbits of a spinless test particle in a 
Schwarzschild background are allowable only for $r>1.5 r_g$ ($r$ is the radial coordinate and $r_g$ is the horizon radius) and the highly relativistic circular orbits exist only for $r=1.5 r_g(1+\delta)$, where $0<\delta \ll 1$
\cite{Misner, Chandra}. The situation with possible circular orbits of a spinning test particle in Schwarzschild's background is another: the space region of existence of the relevant highly relativistic circular orbits is much wider  \cite{Pl05, Pl12}. In particular, it is shown that due to the significant repulsive action of the spin-gravity coupling the highly relativistic circular orbits of a spinning test particle are possible for $r\leq 1.5 r_g$. It means that the corresponding solutions of the MP equations differ essentially from the solutions of the geodesic equations and the worldlines and trajectories of the spinning and spinless particles which start with the same initial values of the coordinates and velocity are not close. In addition, in this section we compare the values of the energy and angular momentum of the corresponding spinning and spinless particles. Like the geodesic equations, the MP equations in Schwarzschild's metric have the constants of motion: the energy $E$ and the angular momentum $J$. 

We take into account the MP equations in the form  \cite{Mathis, Papa}
\begin{equation}\label{1}
\frac D {ds} \left(mu^\lambda + u_\mu\frac {DS^{\lambda\mu}}
{ds}\right)= -\frac {1} {2} u^\pi S^{\rho\sigma}
R^{\lambda}_{~\pi\rho\sigma},
\end{equation}
\begin{equation}\label{2}
\frac {DS^{\mu\nu}} {ds} + u^\mu u_\sigma \frac {DS^{\nu\sigma}}
{ds} - u^\nu u_\sigma \frac {DS^{\mu\sigma}} {ds} = 0,
\end{equation}
where $u^\lambda\equiv dx^\lambda/ds$ is the particle's 4-velocity,
$S^{\mu\nu}$ is the tensor of spin, $m$ and $D/ds$ are,
respectively, the mass and the covariant derivative along $u^\lambda$ and $R^{\lambda}_{~\pi\rho\sigma}$ is
the Riemann curvature tensor (units $c=G=1$ are used). Here, and in
the following, Latin indices run 1, 2, 3 and Greek indices 1, 2, 3,
4; the signature of the metric (--,--,--,+) is chosen.

As usual, these equations are considered with some supplementary condition and most often the Mathisson-Pirani condition \cite{Mathis, Pirani-2}
\begin{equation}\label{3}
S^{\lambda\nu} u_\nu = 0
\end{equation}
or Tulczyjew-Dixon one \cite{Tul, Dixon}
\begin{equation}\label{4}
S^{\lambda\nu} P_\nu = 0
\end{equation}
are used, where
\begin{equation}\label{5}
P^\nu = mu^\nu + u_\lambda\frac {DS^{\nu\lambda}}{ds}
\end{equation}
is the particle 4-momentum. Both at  (\ref{3}) and (\ref{4}), the constant of motion of the MP equations is
\begin{equation}\label{6} S_0^2=\frac12
S_{\mu\nu}S^{\mu\nu},
\end{equation}
where $|S_0|$ is the absolute value of spin. 

In different contexts the MP equations are taken into account in many recent papers \cite{Bini, Obukh, Silenko, Hack, Ram, Singh, Kunst, Jefr, Mohs, Ambr, Ambr2, Kunst2, Costa2, Lanza, Silenko2}.

In \cite{Pl05, Pl12} equations (\ref{1}) and (\ref{2}) are considered in Schwarzschild's metric, using the standard coordinates $x^1=r,$ $x^2=\theta,$ $x^3=\varphi,$ $x^4=t$, to describe the highly relativistic circular orbits of a spinning particle in the plane $\theta=\pi/2$. In these coordinates the constants of the particle's energy and angular momentum are
\begin{equation}\label{7}
E=mu_4+g_{44}u_{\mu}\frac
{DS^{4\mu}} {ds}+\frac{1}{2}S^{\mu 4}g_{44,\mu},
\end{equation}
\begin{equation}\label{8}
J=-mu_3-g_{33}u_{\mu}\frac {DS^{3\mu}} {ds}-\frac{1}{2}S^{\mu 3}g_{33,\mu}.
\end{equation}
In the following we shall use the dimensionless quantities $y_i$ connected with the particle's coordinates and velocity
\begin{equation}\label{9}
\quad y_1=\frac{r}{M},\quad y_2=\theta,\quad y_3=\varphi, \quad
y_4=\frac{t}{M},
\end{equation}
\begin{equation}\label{10}
y_5=u^1,\quad y_6=Mu^2,\quad y_7=Mu^3,\quad y_8=u^4,
\end{equation}
where $M$ is the Schwarzschild mass.
Then the equations which determine the region of existence of the circular orbits of a spinning particle in Schwarzschild's background and the dependence of the particle's angular velocity on the radial coordinate can be written as \cite{Pl12}
$$
y_7^3(y_1-3)^2y_8y_1^{-1}\varepsilon_0-y_7^2(y_1-3)
$$
\begin{equation}\label{11}
+y_7(2y_1-3)\varepsilon_0 y_8y_1^{-3}+y_1^{-2}=0,
\end{equation}
\begin{equation}\label{12}
y_8=\left(1-\frac{2}{y_1}\right)^{-1/2}\sqrt{1+y_1^2 y_7^2},
\end{equation}
where 
\begin{equation}\label{13}
\varepsilon_0\equiv \frac{|S_0|}{mM}\ll 1.
\end{equation}
(Equation (\ref{11}) follows directly from the MP equations (\ref{1}) and (\ref{2}) at condition (\ref{3}) for the Schwarzschild metric when the spinning particle is moving in the plane $\theta=\pi/2$ with the spin orthogonal to this plane, and equation (\ref{12}) is a simple consequence of the condition $u_\mu u^\mu = 1$.) Figures 3--5 in \cite{Pl12} illustrate the dependence of the Lorentz $\gamma$-factor on $r$ for the orbital velocity which is necessary for the particle motions on the circular orbits with $r=const$ in the region $2M<r<3M(1+\delta)$. It is noted in \cite{Pl12} that all orbits in Figs. 3--5 are possible due to the significant repulsive action of the spin-gravity coupling. 

Let us compare  the values of the energy and angular momentum for the spinning and spinless particles which begin to move with $r$ from the region $2M<r<3M(1+\delta)$. We consider the case when a spinning particle is moving on the circular orbits, as in the pointed out above situations from  \cite{Pl12}, and a spinless particle begins  to move with the same initial velocity. There are expressions for $E$ and $J$ according to 
(\ref{7}) and (\ref{8}) in notations (\ref{9}) and (\ref{10}):
\begin{equation}\label{14}
E=m\left(1-\frac{2}{y_1}\right)y_8 - m\varepsilon_0 y_1(y_1-3)y_7^3,
\end{equation}
\begin{equation}\label{15}
J=mMy_1^2 y_7 - mM\varepsilon_0\left(1-\frac{2}{y_1}\right)\left(1-\frac{3}{y_1}\right)
y_8^3.
\end{equation}
Narurally, at $\varepsilon_0=0$ from (\ref{14}) and (\ref{15}) the corresponding expressions follow for the spinless particle.

 Using the corresponding solutions of Eqs. (\ref{11}) and (\ref{12}) in (\ref{14}) and (\ref{15}) we obtain the graphs 
which present the dependence of the particle's energy and angular momentum on $r$ in different subregions of the region $2M<r<3M(1+\delta)$. Figures 1 and 2 show the subregions where the energy and angular momentum of a spinning particle significantly differ from the energy and angular momentum of a corresponding spinless particle and where these values are close. (As in
\cite{Pl12}, here we put $S_2\equiv S_\theta>0$, then $d\varphi/ds<0$; $\varepsilon_0=10^{-2}$). Note that according to  \cite{Pl12} the orbits with $r$ larger then $\approx 3.01$ are not highly relativistic and here $\gamma$-factor is close to 1.

\begin{figure}
\centering
\includegraphics[width=4cm]{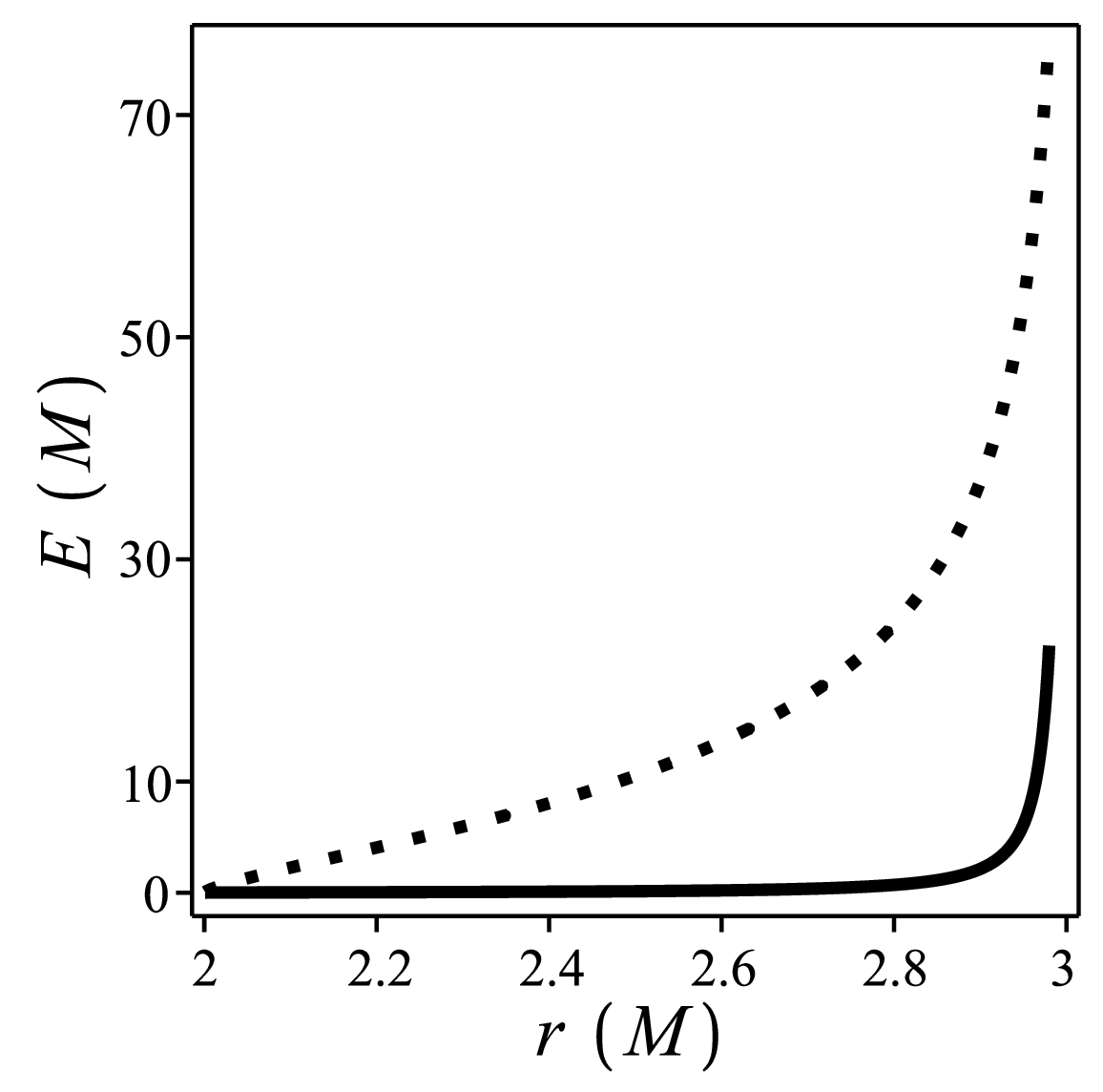}\ \ \
\includegraphics[width=4cm]{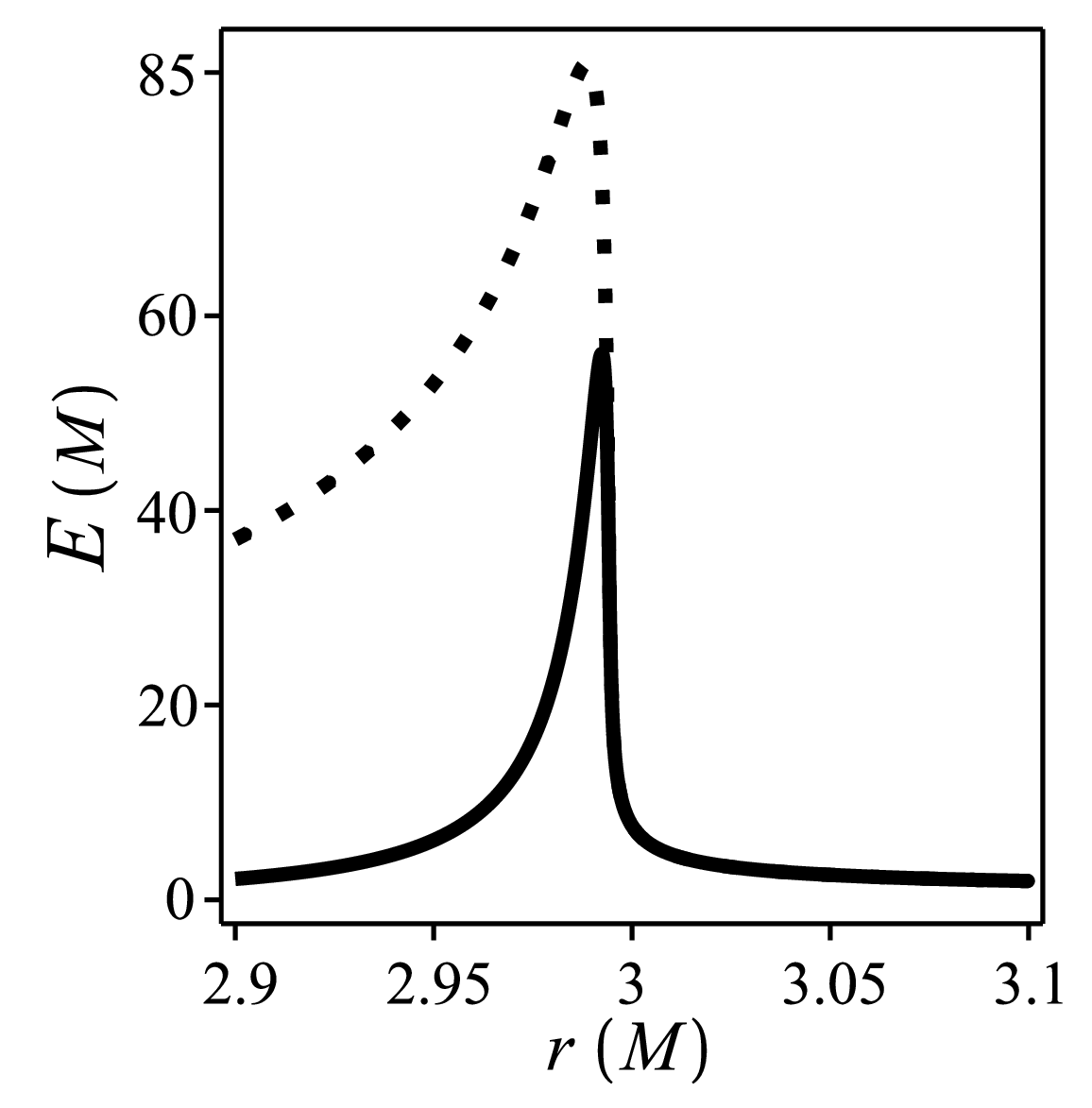}\ \ \
\includegraphics[width=4cm]{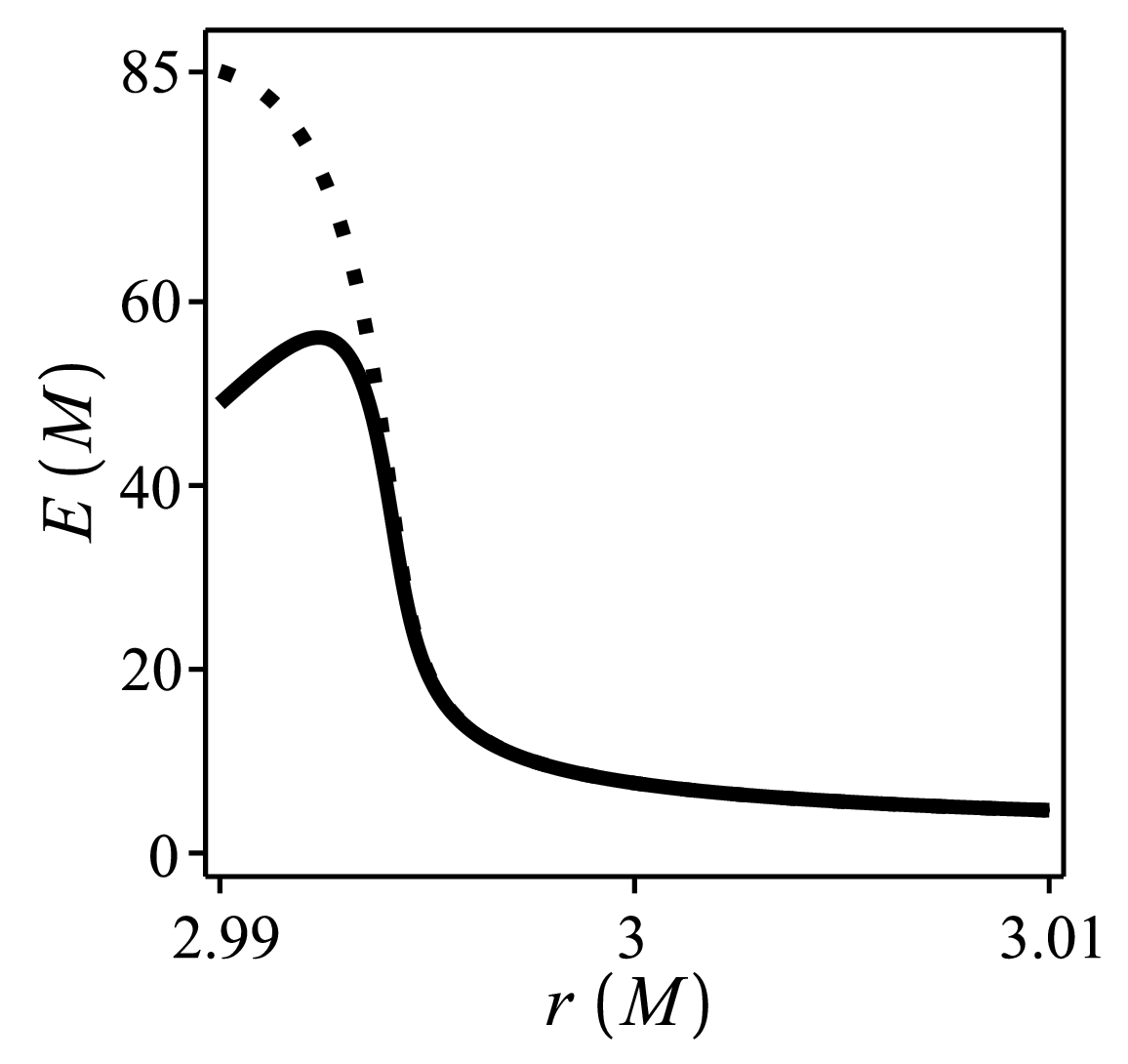}

\caption{\label{1} Energy vs radial coordinate for the circular orbits of a spinning particle (solid line) and for the geodesic motions (dotted line) with the same initial velocity. The three pictures correspond to the different intervals and scaling by $r$. }
\end{figure}

\begin{figure}
\centering
\includegraphics[width=4cm]{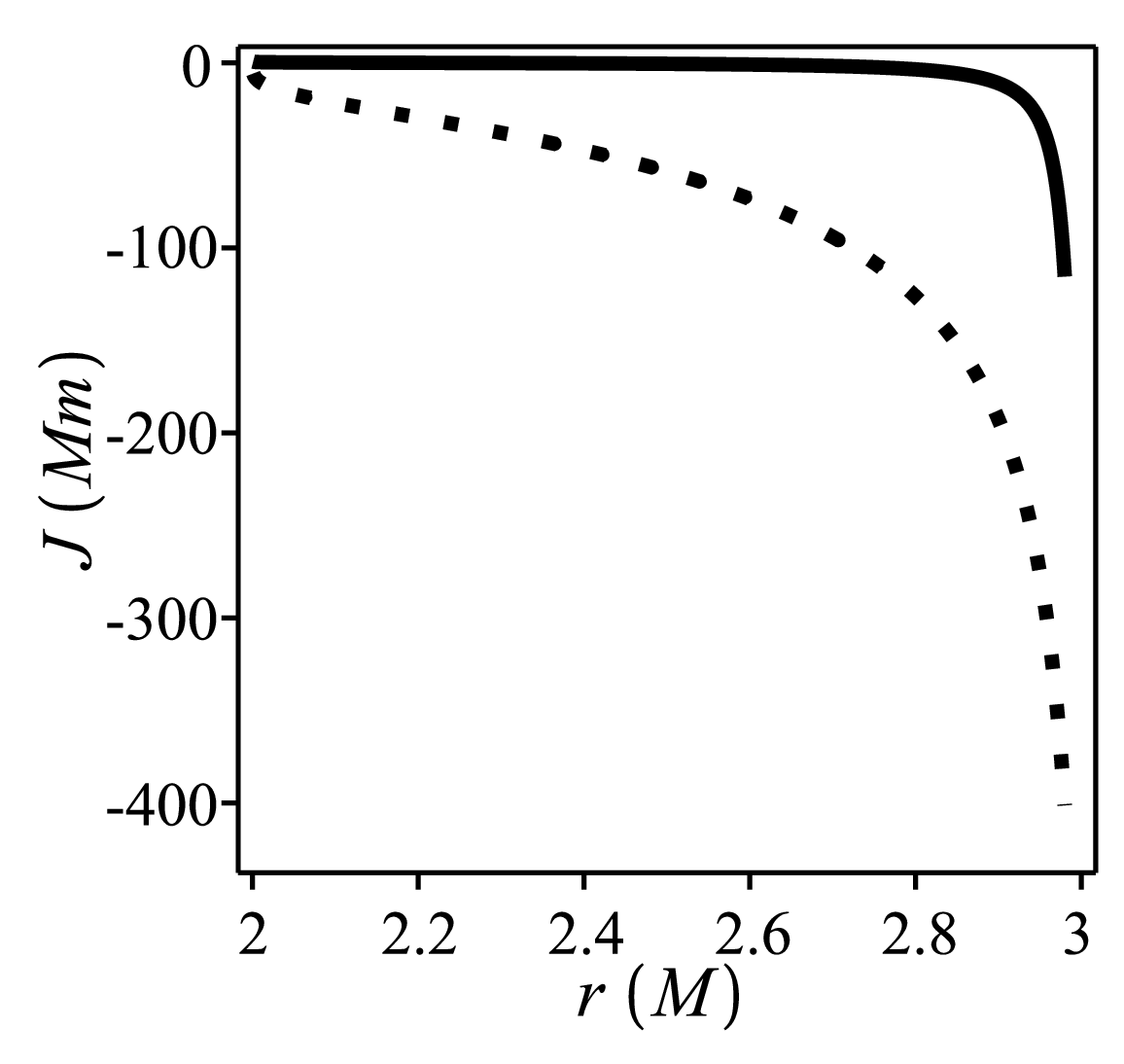}\ \ \
\includegraphics[width=4cm]{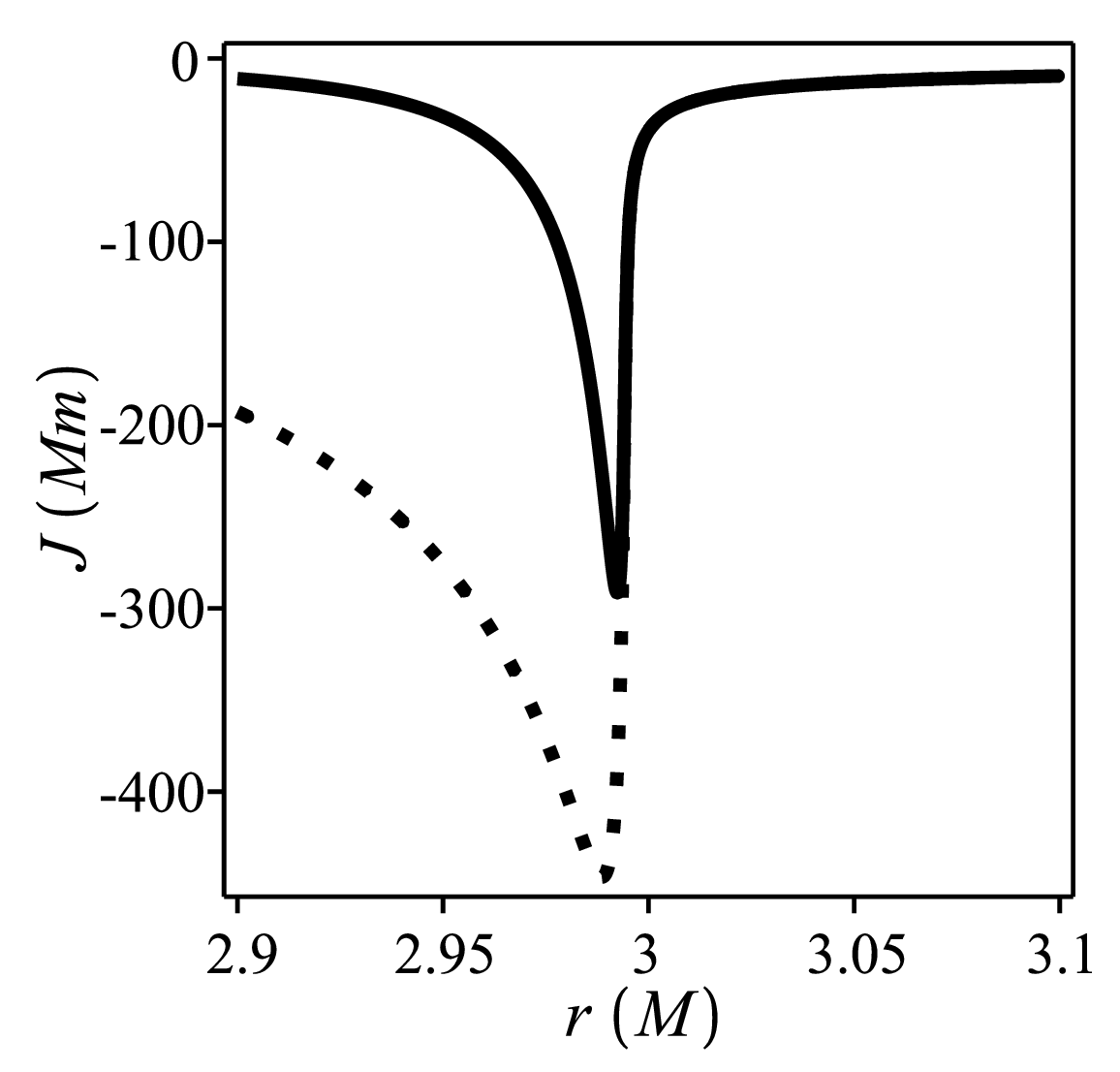}\ \ \
\includegraphics[width=4cm]{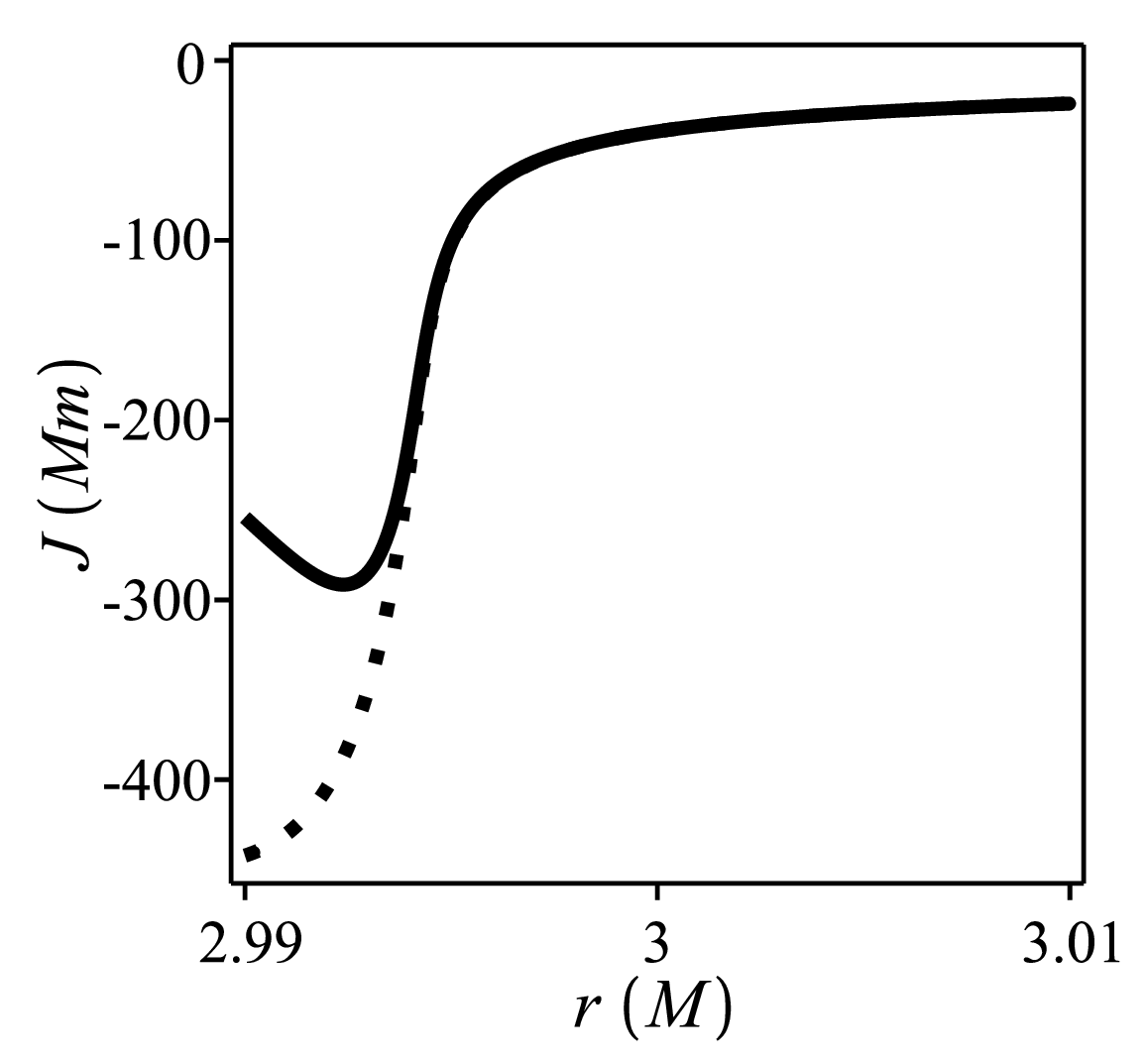}

\caption{\label{2} Angular momentum vs radial coordinate for the circular orbits of a spinning particle (solid line) and for the geodesic motions (dotted line) with the same initial velocity. The three pictures correspond to the different intervals and scaling by $r$.}
\end{figure}

Among all highly relativistic circular orbits of a spinning particle in Schwarzschild's background the orbit with $r=3M=1.5r_g$ has a specific feature: the solution which describes this orbit is the same for the exact MP equations and for their linear spin approximation, and is common for conditions (\ref{3}) and (\ref{4}). Other highly relativistic circular orbits do not have this property. It is pointed out in \cite{Pl12, Pl-Sp, Pl16} that, in general, for the correct description of the highly relativistic orbits of a spinning particle in Schwarzschild's background condition (\ref{3}) is more appropriate.

According to \cite{Pl12} the dependence of the $\gamma$-factor on $\varepsilon_0$ for the highly relativistic circular orbits is determined by the value $1/\sqrt{\varepsilon_0}$. The same dependence on $\varepsilon_0$
takes place for $E$ and $J$ on these orbits.

\section{Beyond the circular orbits}

In addition to the results on the properties of the highly relativistic circular orbits, important information concerning the possibilities of the strong repulsive action on a spinning particle follows from the shape of the highly
relativistic noncircular orbits. We begin from the orbits which start with $r=3M$ and correspond to different values of the particle's orbital velocity.

To describe most general motions of a spinning particle (without restrictions on its velocity and spin orientation) in Schwarzschild's and Kerr's backgrounds by the exact MP equations at condition (\ref{3}), the representation of these equations was developed using the integrals of energy and angular momentum \cite{Pl11, Pl-Sp}. In the more simple particular case of the equatorial noncircular motions of a spinning particle in  Schwarzschild's background the corresponding equations can be written as \cite{Pl08}
 $$
\dot{y_5}=\frac{y_5^2}{y_1}+y_1\left(1-\frac{3}{y_1}\right)\left(2y_7^2+\frac{1}{y_1^2}\right)
-\frac{\hat E}{\varepsilon_0}y_7y_1
$$
\begin{equation}\label{16}
+\frac{\hat{J}}{\varepsilon_0y_1}\left[y_5^2+\left(1-\frac{2}{y_1}\right)(1+y_7^2y_1^2)\right]^{1/2},
\end{equation}
$$
\dot{y_7}=-\frac{y_5y_7}{y_1}+y_1\frac{y_7^2+1/y_1^2}{y_5}\left(y_7-\frac{3y_7}{y_1}-\frac{\hat E}{\varepsilon_0}\right)
$$
\begin{equation}\label{17}
+\frac{1}{y_1 y_5\varepsilon_0}(1+\hat{J}y_7)\left[y_5^2+\left(1-\frac{2}{y_1}\right)(1+y_7^2y_1^2)\right]^{1/2},
\end{equation}
\begin{equation}\label{18}
\dot y_1=y_5, \quad
\dot y_3=y_7,
\end{equation}
where
\begin{equation}\label{19}
\hat E\equiv \frac{E}{m}, \quad  \hat J\equiv \frac{J}{mM},
\end{equation}
and a dot denotes the usual derivative with respect to $x\equiv s/m$.

By choosing different values of $\hat E$ and $\hat J$ for the fixed initial values of $y_i$ one can describe the motions of different centers of mass of a spinning particle \cite{Pl11}. Among the set of the pairs $\hat E$ and $\hat J$ there is the single pair corresponding to the proper center of mass.
(It is known that in Minkowski's spacetime the exact MP equations at condition (\ref{3}) have, in addition to usual solutions describing the straight worldlines, a set of solutions describing oscillatory (helical) worldlines \cite{Mathis2, Weyss}. This situation was interpreted in \cite{Mol1} where it was pointed out that in relativity the position of the center of mass of a rotating body depends on the frame of reference, and 
condition (\ref{3}) is common for the so-called proper and nonproper centers of mass \cite{Mol2}; more detailed analysis can be found in \cite{Costa2, Costa}.)

Concerning the highly relativistic circular orbits of a spinning particle with
$r=3M$ (in notations (\ref{9}) it means $y_1=3$) we note that according to 
(\ref{11}) and (\ref{12}) the value of $y_7$ on this orbit is
\begin{equation}\label{20}
y_7=-\frac{3^{-3/4}}{\sqrt{\varepsilon_0}}(1+O(\varepsilon_0)). 
\end{equation}
Then by expressions (\ref{14}) and (\ref{15}) in the main approximation we have
\begin{equation}\label{21}
\hat E=\frac{3^{-1/4}}{\sqrt{\varepsilon_0}}, \quad
\hat J=-\frac{3^{5/4}}{\sqrt{\varepsilon_0}}.
\end{equation}

Let us consider the highly relativistic noncircular motions of a spinning particle which starts from $y_1=3$ with the initial values of $y_5\ne 0$ and
with $y_7$ which differs from (\ref{20}). For this purpose we integrate equations (\ref{16})--(\ref{18}). The values of $\hat E$ and $\hat J$ which correspond to the motions of the proper center of mass can be found using search computer. As typical, in Fig. 3 we show the results concerning the shape of the spinning particle trajectories at $\varepsilon_0=10^{-2}$ with the fixed initial values $y_1(0)=3, \quad y_5(0)=-2.5\times 10^{-2}$ and different values of $y_7$. The solid line corresponds to $y_7(0)\approx -4.39$: this value is determined by (\ref{20}) with $\varepsilon_0=10^{-2}$. The dashed line, long dash line, and dash-dotted
lines describe the cases when $y_7(0)$ is equal approximately to $-4.39$
multiplied by 2, 4 and 6 correspondingly. In all cases the particle starts clockwise from the position $r=3M$ and $\varphi=0$, in the polar coordinates. According to the solid line the spinning particle with the corresponding initial values of $y_7$ falls on Schwarzschild's horizon surface as well as the spinless particle which begins to move with the same initial conditions (for comparison the dot line in Fig. 3 illustrates the trajectory of this spinless particle). The three other curves (dashed, long dash, and dash-dotted lines) show that the spinning particle with the corresponding initial values of $y_7$ goes away from the Schwarzschild source, whereas it is known that by the properties of the geodesic lines in Schwarzschild's metric the spinless particle in all these cases falls on the horizon surface similarly as the dot line in Fig. 3. So, Fig.3 illustrates how the spin-gravity action on the trajectory of the spinning particle increases with its orbital velocity $y_7$.
\begin{figure}
\centering
\includegraphics[width=6cm]{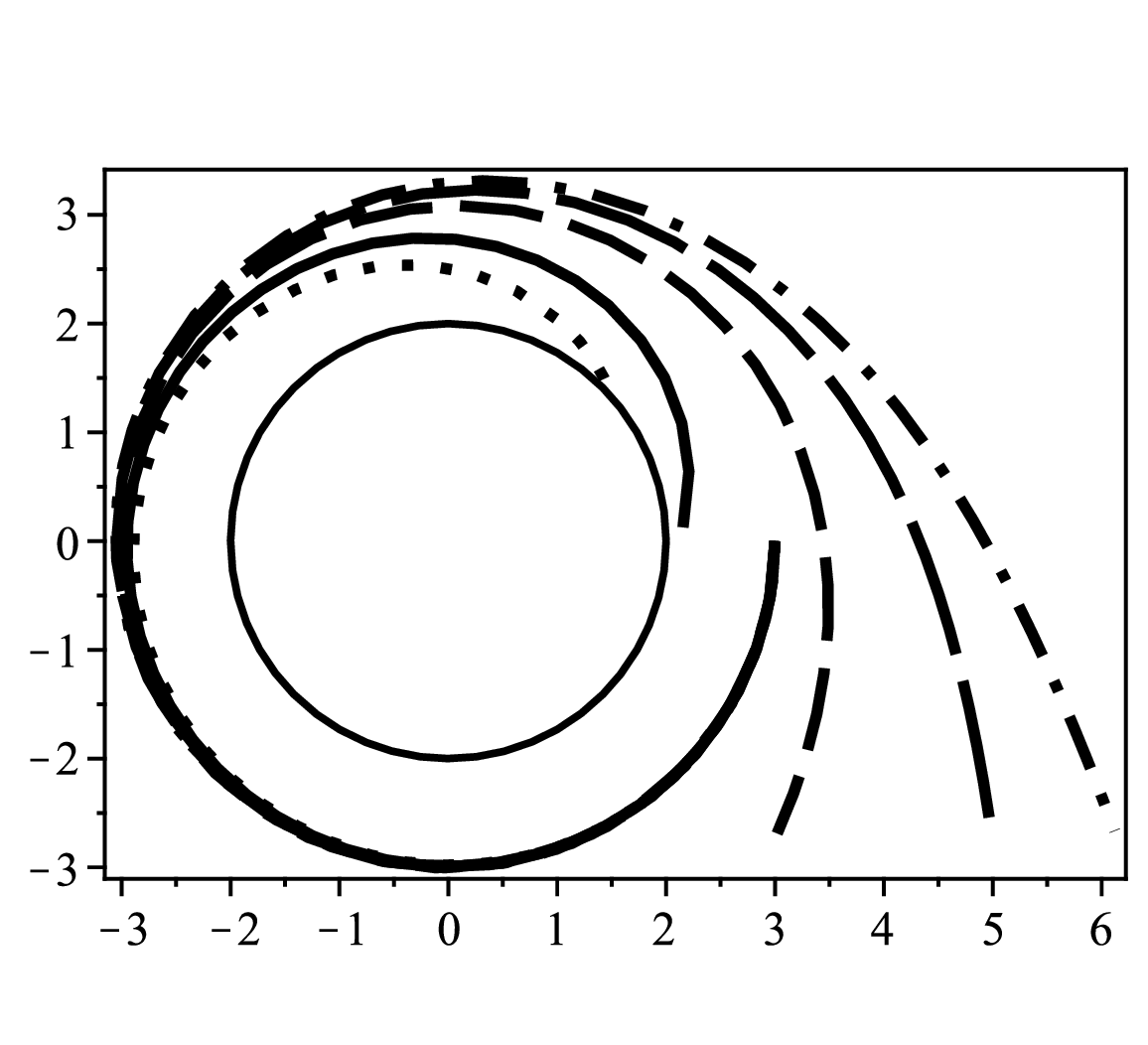}\ \ \
\caption{\label{3}Noncircular trajectories of the spinning particle at different initial values of $y_7$. The circle $y_1=2$ corresponds to the horizon line. }
\end{figure}

Note that the highly relativistic circular orbits of a spinning particle in Schwarzschild's background exist beyond the small neighborhood of the value $y_1=3$ as well, for $2<y_1<3$. Then the necessary value of $y_7$ is determined by
$$
y_7=-\frac{1}{\sqrt{\varepsilon_0 y_1}}
\left(1-\frac{2}{y_1}\right)^{1/4}
$$
\begin{equation}\label{22}
\times \left(\frac{3}{y_1}-1\right)^{-1/2}(1+O(\varepsilon_0)).
\end{equation}
Similarly to (\ref{20}) expression (\ref{22}) is proportional to $1/\sqrt{\varepsilon_0}$. The expressions for $\hat E$ and $\hat J$ which
correspond to (\ref{22}) are
$$
\hat E=\frac{\sqrt{\varepsilon_0}}{\sqrt{y_1}}
\left(1-\frac{2}{y_1}\right)^{1/4}\left(\frac{3}{y_1}-1\right)^{-3/2}
$$
\begin{equation}\label{23}
\times \left(1-\frac{3}{y_1}+\frac{3}{y_1^2}\right),
\end{equation}
$$
\hat J= \sqrt{\varepsilon_0
y_1}\left(1-\frac{2}{y_1}\right)^{-1/4}\left(\frac{3}{y_1}-1\right)^{-3/2}
$$
\begin{equation}\label{24}
\times\left(1-\frac{9}{y_1}+\frac{15}{y_1^2}\right).
\end{equation}
That is, in contrast to (\ref{21}), here both $\hat E$ and $\hat J$ are proportional to $\sqrt{\varepsilon_0}$.

To describe noncircular highly relativistic orbits of a spinning particle which starts with $2<y_1(0)<3$ one can use search computer of such values $\hat E$ and $\hat J$ which pick out the corresponding motions of the particle's proper center of mass, similarly as it was pointed out above for the motion with $y_1(0)=3$. Naturally, if some noncircular orbits are closer, in a certain sense, to the circular orbits with (\ref{22}), the corresponding values $\hat E$ and $\hat J$ are close to (\ref{23}) and (\ref{24}). In general, it is useful to have some analytical estimations for the necessary 
$\hat E$ and $\hat J$ in any concrete case of the spinning particle motion.

\section{On the values of $\hat E$ and $\hat J$ for the proper center of mass}

The procedure of finding the $\hat E$ and $\hat J$ for the proper center of mass of a spinning particle moving in Schwarzschild's background which is based on the consideration of the MP 
equations in the linear approximation by the small displacement of the  values $y_1, y_5, y_7 $ from their initial values $y_1(0), y_5(0), y_7(0)$ is
presented in \cite{Pl08}. It is important in this approach that $\hat E$ and $\hat J$ are the constant of motion, that is their values are the same for the all time of the particle's motion. In notations
$$
\xi_{1} \equiv \frac{y_5 - y_5(0)}{y_5(0)}, \quad \xi _{2}
\equiv \frac{y_7 - y_7(0)}{y_7(0)},
$$
\begin{equation}\label{25}
 \xi _{3} \equiv \frac{y_1 -y_1(0)}{y_1(0)}.
\end{equation}
it follows from  (\ref{16}) and (\ref{17}) in the linear in $\xi_i$ approximation 
$$
\dot{\xi_1} = (a_{10} + a_{11}\hat{J}\varepsilon_0^{-1} + a_{12}\hat{E}\varepsilon_0^{-1})\xi_{1}
+ (a_{20} + a_{21}\hat{J}\varepsilon_0^{-1}
$$
$$
 + a_{22} \hat{E}\varepsilon_0^{-1})\xi_{2}+ ( a_{30} + a_{31} \hat J \varepsilon_0^{-1} + a_{32} \hat E \varepsilon_0^{-1})\xi_{3}
$$
\begin{equation}
\label{26}
+ a_{00} + a_{01} \hat J \varepsilon_0^{-1} + a_{02} \hat E \varepsilon_0^{-1} ,
\end{equation}
$$
\dot \xi _2 = ( b_{10} + b_{11} \hat J \varepsilon_0^{-1} + b_{12} \hat E \varepsilon_0^{-1}
+ b_{13} \varepsilon_0^{-1})\xi _{1} 
$$
$$
+ ( b_{20} +
b_{21} \hat {J} \varepsilon_0^{-1} + b_{22} \hat E \varepsilon_0^{-1} + b_{23} \varepsilon_0^{-1})\xi_{2}
$$
$$
+ ( b_{30} + b_{31} \hat J \varepsilon_0^{-1} + b_{32} \hat E \varepsilon_0^{-1} + b_{33}
\varepsilon_0^{-1})\xi _{1}
$$
\begin{equation}\label{27} 
 + b_{00} + b_{01} \hat J \varepsilon_0^{-1} +
b_{02} \hat E \varepsilon_0^{-1} + b_{03} \varepsilon_0^{-1},
\end{equation}
\begin{equation}
\label{28}
 \dot\xi_3 = c_{10}\xi_{1} + c_{00},
\end{equation}
where the coefficients $a, b, c$ with the corresponding indexes are expressed through
$y_1(0), y_5(0), y_7(0)$ as follows:
$$
a_{10}=2y_1^{-1}y_5, \quad a_{11}=y_1^{-1}y_5N,
$$
$$
a_{12}=0, \quad a_{20}=4y_5^{-1} y_7^2 (y_1-3),
$$
$$
a_{21}=(y_1-2)y_5^{-1}y_7^2N, \quad a_{22}=-y_1y_5^{-1}y_7,
$$
$$
a_{30}=y_1^{-2}y_5^{-1}[y_1(2y_1^2y_7^2-y_5^2-1)+6],
$$
$$
a_{31}=y_1^{-1}y_5^{-1}N(-1-y_5^2+y_1y_7^2+3y_1^{-1}),
$$
$$
a_{32}=-y_1y_5^{-1}y_7,
$$
$$
a_{00}=y_1^{-1}y_5+(y_1-3)y_5^{-1}(y_1^{-2}+2y_7^2),
$$
$$
a_{01}=y_1^{-1}y_5^{-1}N^{-1}, \quad a_{02}=-y_1y_5^{-1}y_7;
$$
$$
b_{10}=-y_1^{-1}y_5-y_1^{-2}y_5^{-1}(y_1-3)(1+y_1^2y_7^2),
$$
$$
b_{11}=-y_1^{-1}y_5^{-1}N(1+y_1^2y_7^2)(1-2y_1^{-1}),
$$
$$
b_{12}=y_1y_5^{-1}y_7{-1}(y_1^{-2}+y_7^2),
$$
$$
b_{13}=-y_1^{-1}y_5^{-1}y_7^{-1}N(1+y_1^2y_7^2)(1-2y_1^{-1}),
$$
$$
b_{20}=-y_1^{-1}y_5+y_1^{-2}y_5^{-1}(y_1-3)(1+3y_1^2y_7^2),
$$
$$
b_{21}=y_1^{-1}y_5^{-1}N[y_5^2+(1-2y_1^{-1})(1+2y_1^2y_7^2),
$$
$$
b_{22}=-2y_1y_5^{-1}y_7, \quad b_{23}=y_5^{-1}y_7N(y_1-2),
$$
$$
b_{30}=y_1^{-2}y_5^{-1}[6+y_1(y_5^2+y_1^2y_7^2-1)],
$$
$$
b_{31}=y_1^{-1}y_5^{-1}N(-1-y_5^2+3y_1^{-1}+y_1y_7^2),
$$
$$
b_{32}=y_1^{-1}y_5^{-1}y_7^{-1}(1-y_1^2y_7^2), \quad b_{33}=y_7^{-1}b_{31},
$$
$$
b_{00}=-y_1^{-1}y_5+y_1^{-2}y_5^{-1}(y_1-3)(1+y_1^2y_7^2),
$$
$$
b_{01}=y_1^{-1}y_5^{-1}N^{-1}, \quad b_{02}=-y_1^{-1}y_5^{-1}y_7^{-1}(1+y_1^2y_7^2),
$$
$$
b_{03}=y_1^{-1}y_5^{-1}y_7^{-1}N^{-1};
$$
\begin{equation}\label{29}
c_{00}=c_{10}=y_1^{-1}y_5,
\end{equation}
where 
$$
N=[y_5^2+(1-2y_1^{-1})(1+y_1^2y_7^2)]^{-1/2}
$$
(for brevity we omit "(0)" near the initial values of $y_i$ in (\ref{29}) and in the following).

According to the known result  of the differential equations theory, the general solution of linear
equations (\ref{26})--(\ref{28}) is determined by the combination of $e^{\lambda_ix}$
(i=1,2,3), where $\lambda_i$ are the solutions of the third-order algebraic equation
\begin{equation}
\label{30} \lambda ^{3} + C_{2} \lambda ^{2} + C_{1} \lambda + C_{0} = 0.
\end{equation}
Here the coefficients $C_j$ ($j=0,1,2$) can be expressed through $a, b, c$ and depend
both on $y_1(0), y_5(0), y_7(0), \varepsilon_0$ and on the parameters $\hat E$ and  $\hat J$. For example, the corresponding expressions for $C_2$ and $C_1$ are
$$
C_2=-a_{10}-b_{20}-\hat J \varepsilon_0^{-1}(a_{11}+b_{21})
$$
\begin{equation}\label{31}
 -\hat E \varepsilon_0^{-1}(a_{12}+b_{22})-b_{23}\varepsilon_0^{-1},
\end{equation}
$$
C_1=(a_{10}+a_{11}\hat J \varepsilon_0^{-1}+a_{12}\hat E \varepsilon_0^{-1})(b_{20}+b_{21}\hat J\varepsilon_0^{-1}
$$
$$
+b_{22}\hat E \varepsilon_0^{-1}+b_{23}\varepsilon_0^{-1})-c_{10}(a_{30}+a_{31}\hat J \varepsilon_0^{-1}+a_{2}\hat E \varepsilon_0^{-1})
$$
$$
-(a_{20}+a_{21}\hat J \varepsilon_0^{-1}+a_{22}\hat E \varepsilon_0^{-1})
$$
\begin{equation}\label{32}
\times (b_{10}+b_{11}\hat J \varepsilon_0^{-1})+b_{12}\hat E\varepsilon_0^{-1}).
\end{equation}

Let us consider (\ref{31}) and (\ref{32}) for the concrete cases of the particle's highly relativistic noncircular motions when its 4-velocity is determined by the relationships
\begin{equation}\label{33}
y_5=\frac{p}{\sqrt{\varepsilon_0}}, \quad
y_7=\frac{k}{\sqrt{\varepsilon_0}},
\end{equation}
where the parameters $p$ and $k$ satisfy the conditions $p^2/\varepsilon_0\gg 1$ and $k^2/\varepsilon_0\gg 1$. That is, similarly to the case of the circular orbits with (\ref{22}), according to (\ref{33}) the
particles 4-velocity is proportional to $1/\sqrt{\varepsilon_0}$. Our task is to find such values $\hat E$ and  $\hat J$ which at the fixed initial values $y_1, y_5$ and $y_7$ determine just the motion of the proper center of mass. Using some analogy with the highly relativistic circular orbits when the necessary values $\hat E$ and  $\hat J$ are determined by (\ref{23}) and (\ref{24}), here we search for the corresponding values in the form
\begin{equation}\label{34}
\hat E=k_1\sqrt{\varepsilon_0}, \quad
\hat J=k_2\sqrt{\varepsilon_0},
\end{equation}
where $k_1$ and $k_2$ are some parameters which we have to find. For this purpose we take into account the known expressions for the roots of the third order algebraic Eq.  (\ref{30}) through the values $C_2$, $C_1$
and $C_0$. It follows from these expressions that the values of the three roots $\lambda_1$,  $\lambda_2$ and  $\lambda_3$ significantly depend on $\varepsilon_0$. In general, the expressions for these roots contain the large terms which are proportional to $1/\varepsilon_0$. Just these terms determine the high frequency oscillatory solutions as well as the solutions which are proportional to the exponent with the large absolute values of the 
real index of this exponent. Such all solutions do not describe the motions of the particle's proper center of mass in which we are interested. Therefore, to choice the necessary solutions we take into account the partial solutions of Eqs. (\ref{26})--(\ref{28}) for which the corresponding expressions of 
$\lambda_1$,  $\lambda_2$ and  $\lambda_3$ do not contain the large terms of the order $1/\varepsilon_0$. It is not difficult to check that the possible approximated approach consists in putting zero the coefficients near the terms with $1/\varepsilon_0$ in the expressions for $C_2$ and $C_1$. Then we obtain the two linear algebraic equations for $k_1$ and $k_2$ which determine the necessary values of these parameters:
\begin{equation}\label{35}
k_1=\frac{k_2n}{y_1^2k}+\frac{p^2}{2y_1^2k}+\frac{3k}{2y_1}(y_1-3)+
\frac{y_1-2}{2y_1n},
\end{equation}
\begin{equation}\label{36}
k_2=\frac{A}{B},
\end{equation}
where
$$
A=2.75p^2n-k^2y_1n(y_1-9)+1.25k^4y_1^2np^{-2}(y_1-3)^2
$$
$$
-1.5k^3y_1^2(y_1-2)(y_1-3)p^{-2}+0.25k^2y_1^2(y_1-2)^2p^{-2}n^{-1},
$$
$$
B=k^2y_1(y_1-3),
$$
$$
n=\sqrt{p^2+k^2y_1(y_1-2)}, \quad y_1\ne 3.
$$

Let us apply relationships (\ref{34})--(\ref{36}) in Eqs. (\ref{16})--(\ref{18}) for the concrete motions of a highly relativistic spinning particle with the initial value $y_1(0)=2.5$. Figure 4 shows a case when $y_5(0)=3.6$, $y_7(0)=-9.2$ and $\varepsilon_0=10^{-2}$ (the dot lines in Fig. 4 correspond to the motion of a spinless particle with the same initial values $y_1(0)$,  $y_5(0)$ and $y_7(0)$). Because expressions (\ref{35}) and (\ref{36}) can be used only for some approximated description of motions of the particle's proper center of mass,
the graphs in Fig. 4 have the oscillatory features. It means that the proper center of mass is moving according to the corresponding middle lines of the graphs in Fig. 4.

Note that both graphs  in Fig. 3 and Fig. 4 correspond to the spinning particle motions under the strong repulsive action of the highly relativistic spin-gravity coupling.
\begin{figure}
\centering
\includegraphics[width=4cm]{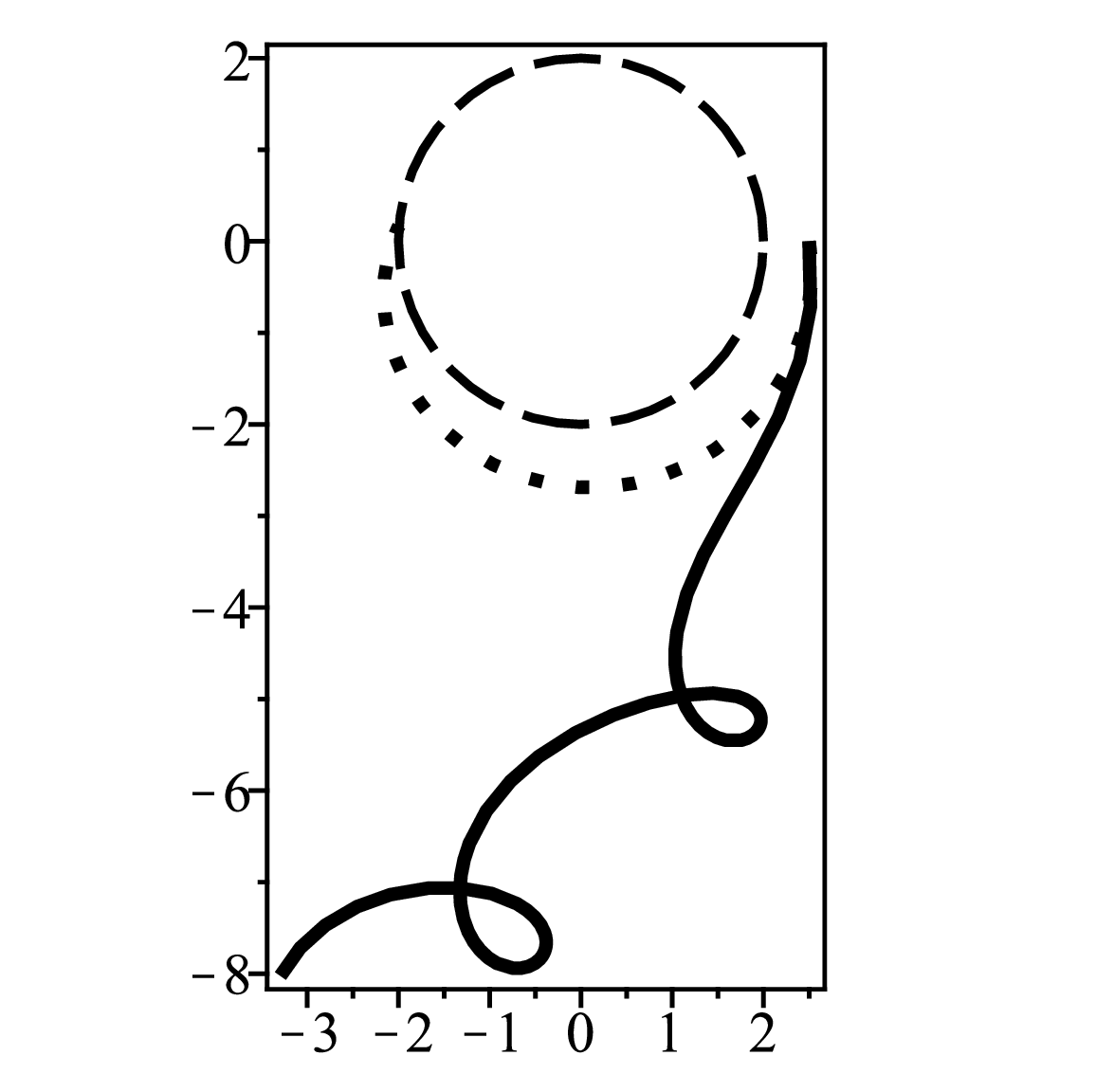}\ \ \
\includegraphics[width=4cm]{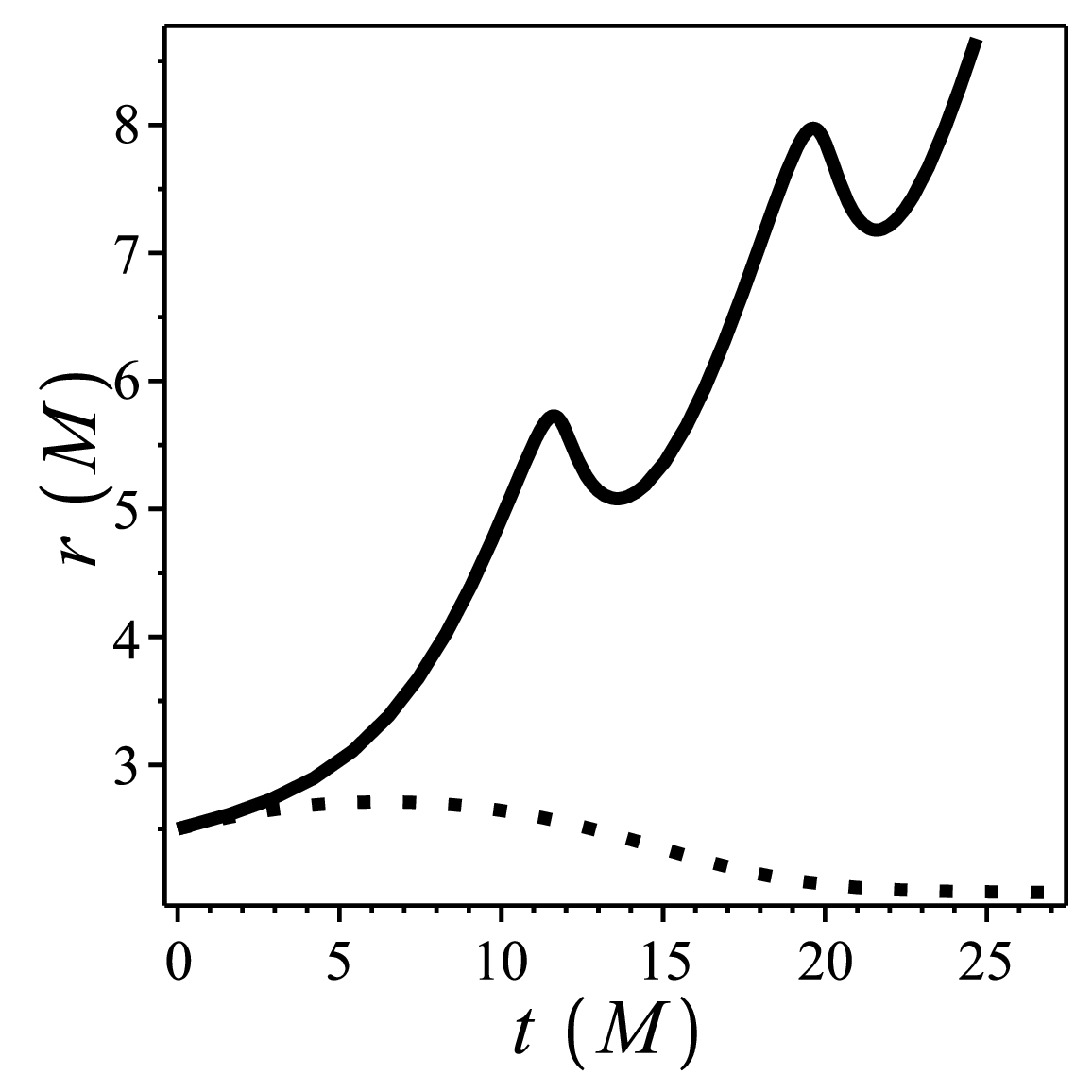}\ \ \

\caption{\label{4} An example of the oscillatory motions in the polar coordinates and in the dependence $r$ vs $t$. }
\end{figure}

\section{Conclusions}

There are significant differences in the spin-gravity coupling for a spinning test particle in  Schwarzschild's background when its velocity (1.) is not very high and (2.) is very close to the speed of light, i.e. when the corresponding relativistic Lorentz factor $\gamma$ is of the order 1 or much greater than 1. Just in the second case it follows from the MP equations that general relativity is both the theory of gravity as a generalization of the Newtonian  description of gravity and, in the certain sense,  predicts the effects of strong antigravity in some extremal situations which is impossible in the Newtonian theory. In this paper we have considered the examples of the highly relativistic motions of a spinning particle which are caused by the strong repulsive action of the spin-gravity coupling in Schwarzschild's background. This action is shown in the form of the spinning particle trajectories as compared to the corresponding geodesic trajectories of a spinless particle (Secs. III and IV). The effects of the significant influence of the highly relativistic spin-gravity coupling on the spinning particle energy and angular momentum are presented in Sec. 2. For further investigations of the concrete types of the highly relativistic motions of a spinning particle in Schwarzschild's background according to the exact MP equations one can use the approach which is described in Sec. IV. 

The question arises concerning possibilities  of the experimental registration of the strong spin-gravity effects. Naturally, the situation with a macroscopic spinning particle (body) moving relative to Schwarzschild's mass with $\gamma\gg 1$ is not realistic. Quite the reverse, the elementary particles which are the active participants of the high energy astrophysical processes have the very large $\gamma$-factor. Which values of $\gamma$ are necessary for the manifestations of the effects of the highly relativistic spin-gravity coupling that are considered in this paper? Note that the main large term which determines these values is equal to $1/\sqrt{\varepsilon_0}$. In the case when $M$ is equal to $10^{6}$ of the Sun's mass, for an electron and a neutrino (with the mass $\approx 0.3 eV$) we have that $1/\sqrt{\varepsilon_0}$ is equal to $9\times 10^{10}$ and $7\times 10^{7}$ respectively. These values are very high and the parts of electrons and neutrinos with the corresponding values of the
$\gamma$-factor near the black holes are low. Nevertheless, one cannot exclude that, for example, the data concerning very high energy neutrinos from the Ice Cube experiment will be useful in the context of the possible registrations of the strong spin-gravity coupling effects.

\end{document}